                        \newif\ifpaper \newif\ifPDF               
                        \newif\ifOUP \newif\ifboyscout            
                        \newif\ifdasbuch \newif\ifarticle         
                        \newif\ifsolutions

                        \newif\ifboyscout                         
                        \newif\ifpreparepdf                       
                        \preparepdftrue 

        \ifboyscout
\documentclass[final,1p,times,showpacs,hyperref]{elsarticle}
        \else
\documentclass[final,1p,times,showpacs,superscriptaddress]{elsarticle}
        \fi

\usepackage[export]{adjustbox} 
\graphicspath{{figs/}{../figs/}}  

\ifpreparepdf
    \usepackage{color} 
    \usepackage[colorlinks]{hyperref} 
\else 
\fi
    \usepackage{graphicx}
    \usepackage{ifthen}

\ifpreparepdf 

\else  

\fi

        \ifboyscout 

   \newcommand{\PC}[2]{$\footnotemark\footnotetext{#1 Predrag: #2}$}

   \newcommand{\DL}[2]{$\footnotemark\footnotetext{#1 Domenico: #2}$}
   \definecolor{darkgreen}{rgb}{0.0, 0.5, 0.0}

   \newcommand{\JMH}[2]{$\footnotemark\footnotetext{#1 Jeffrey: #2}$}
   
        \else  
   \newcommand{\PC}[2]{}{}
   
   \newcommand{\DL}[2]{}{}
   
   \newcommand{\JMH}[2]{}{}

        \fi

\newcommand{\rf}     [1] {~\cite{#1}}
\newcommand{\refref} [1] {ref.~\cite{#1}}

\newcommand{\refeq}  [1] {(\ref{#1})}
\newcommand{\reffig} [1] {figure~\ref{#1}}
\newcommand{\refFig} [1] {Figure~\ref{#1}}
\newcommand{\refsect}[1] {sect.~\ref{#1}}

\newcommand{\beq}{\begin{equation}}

\newcommand{\nnu}{\nonumber}
\newcommand{\eeq}{\end{equation}}
\newcommand{\ee}[1] {\label{#1} \end{equation}}
\newcommand{\bea}{\begin{eqnarray}}

\newcommand{\eea}{\end{eqnarray}}

\newcommand{\ssp}{\ensuremath{x}}    

\newcommand{\msr}{{\rho}}               
\newcommand{\pS}{{\cal M}}          

\newcommand{\diffTen}{\ensuremath{\Delta}}  
\newcommand{\Lnoise}[1]{{\cal L}^{#1}}    
\newcommand{\transp}[1]{{#1}{}^\top}
\newcommand{\matId}{\ensuremath{{\bf 1}}}      

\usepackage{amssymb}

\journal{Communications in Nonlinear Science and Numerical Simulation}

\begin{document}

\begin{frontmatter}



\title{Perturbation theory for the Fokker-Planck operator in chaos}

\author{Jeffrey M. Heninger$^{\mathrm{a}}$}
\author{Domenico Lippolis$^{\mathrm{b},}$\footnote{domenico@ujs.edu.cn}}
\author{Predrag Cvitanovi\'c$^{\mathrm{d}}$}
\address{$^\mathrm{a}$Department of Physics,  University of Texas, Austin, TX, USA}
\address{$^\mathrm{b}$Faculty of Science, Jiangsu University, Zhenjiang 212013, China}
\address{$^\mathrm{d}$Center for Nonlinear Science and School of Physics, Georgia Institute of Technology,
                Atlanta, GA, USA}

\begin{abstract}

The stationary distribution of a fully chaotic system typically exhibits a fractal structure,
which dramatically changes if the dynamical equations are even slightly modified.
Perturbative techniques are not expected to work in this situation. In contrast,
the presence of additive noise smooths out the stationary distribution,
and
perturbation theory becomes applicable.
We show that
a perturbation expansion for the Fokker-Planck evolution operator
yields surprisingly accurate estimates of long-time
averages in an otherwise unlikely scenario.


\end{abstract}


\end{frontmatter}


\section{Introduction}

The divergence of perturbation series due to small denominators in the
vicinity of unstable fixed points, observed by Poincar\'e\rf{poincare},
was historically among the first hints of chaotic dynamics.
This is enough to
tell us that perturbative analysis might not get along with chaos.

Nonetheless, there have been successful attempts to treat instabilities
with perturbation techniques, the first tracing back to
Shimizu\rf{Shim1,Shim2,Shim3}. Shimizu model systems have a parameter
which gradually drives them from periodicity to the onset of chaos. In
that context, fast and slow time scales in the periodic phase allow for
closed-form, asymptotic solutions, which shed light on the transition to
chaos.

The validity of perturbation theory for systems far from equilibrium
is intimately related to the existence of a
linear response to the perturbation\rf{Kubo57},
an issue that has sparked great interest
and some controversy\rf{KampenKubo,SaitoMatsu,Suhl94}
over the years. For maps in particular,
we know that all smooth hyperbolic\rf{Ruelle97} and some partially
hyperbolic diffeomorphisms\rf{Dolgo04} admit a linear response.
The question is instead still open for a variety of physically
interesting models, including the H\'enon family, as well as piecewise
hyperbolic maps\rf{Ruelle09,Baladi_lr}.

The object of our investigation is systems already deep in the
fully chaotic regime, where in general one cannot separate time scales.
We wish to determine whether the dynamics ever
lends itself to a perturbative approach. The issue was notably
addressed more than $20$ years ago by Ershov\rf{Ershov_dilemma}
in the context of one-dimensional maps, with the result that
a perturbation of $O(\epsilon)$ to the original map would
bring about a deviation up to $O(\epsilon\log\epsilon)$
in the response (see \refref{KellerEpsLogEps} for a proof).
Chaos breaks the proportionality
between control parameter and statistical characteristics, disrupting
the very basis of perturbation theory.
Shortly later Ershov\rf{Ershov_contract} established that the Perron-Frobenius
operator becomes continuous in the presence of
additive, uncorrelated noise. The response to
perturbations of a density is then restored to $O(\epsilon)$,
allowing in principle for perturbative calculations.

Proportionality between parameters and observables plays
a major role in chaotic time series analysis\rf{Durb_Koop}, as first pointed out in \refref{XZTang95},
where perturbations may be fruitfully used to test the robustness of the model\rf{Mayb_Amri}.
In particular, dealing with long time series often boils down to building
a Markov chain from the data. This can be expressed as a discretization of the
Perron-Frobenius operator of the dynamics\rf{Froy-2}, whose leading eigenfunction,
or invariant density, is used for the estimation of long time averages\rf{DasBuch}.

We organize the paper as follows:
the \textit{Fokker-Planck evolution operator} for a discrete-time
dynamical system is introduced in \refsect{secFPop}. In \refsect{PT} we
add a small deterministic correction to a weakly noisy map,
and develop a perturbation technique to approximate the corresponding
Fokker-Planck operator. Our goal is to accurately
estimate observables such as the escape rate from a noisy attractor, by
using the stationary distribution (or invariant density, for a closed
system) of the unperturbed noisy system. Direct numerical simulations of
the weakly-noisy Lozi map with a small correction are compared to the
perturbative approach in \refsect{numcs}. Our main result is that the
perturbative estimates are in a close agreement with the outcomes of
direct numerical simulations for a range of perturbations about an order
of magnitude larger than expected. That constitutes quantitative evidence
that the noise restores structural stability of the system, with the
response proportional to the noise amplitude. Conclusions and comments
are given in \refsect{summ}.

\section{The Fokker-Planck operator}
\label{secFPop}
The problem that we are considering involves adding a random variable
to a map in $d$ dimensions that would otherwise exhibit deterministic chaos
(the subscript $n$ represents time iteration):
\beq
	x_{n+1} = f(x_n) + \xi_n
\eeq
The random variables $\xi_n$ are uncorrelated in time, independent, and distributed
according to a Gaussian with a $[d\!\times\!d]$ covariance matrix
\beq
\diffTen(x_n)_{ij} = \langle\xi_n^{(i)}\xi_n^{(j)}\rangle \ \delta_{ij}
\,,
\ee{noise_var}
which we allow to vary in position $x_n$ but not in time.
Time is referenced by $n$. $i$ and $j$ range from 1 to $d$.
Here we shall consider the evolution of densities of trajectories,
according to the Fokker-Planck picture\rf{risken96}.
In discrete time, a distribution moves one time step according to the operator\rf{LipCvi08,CviLip12}
\bea
	\phi_{n+1} (y) &=& \Lnoise{} \phi_n \ (y)
	= \int \phi_n (x) \ \exp\left\{-\frac{1}{2} \transp{[y-f(x)]} \diffTen^{-1}(x) [y-f(x)]\right\} \ [dx]
\label{FPop}\\
{[dx]} &=& \frac{d^d\!x}{|\det(2 \pi\diffTen(x))|^{1/2}}
\nnu
\,.
\eea

The best information we can get about the long time behavior
of the system is statistical. Recalling that we are interested in perturbations
of steady states, the expectation value of any observable $a(x)$ can be found by knowing
the escape rate $\gamma$ and stationary distribution $\rho(x)$
\beq
	\langle a \rangle = \int e^{\gamma} \ \rho(x) \ a(x) \ dx.
\eeq

The escape rate and the  stationary distribution
are respectively the logarithm of the maximal modulus (leading) eigenvalue and
the leading eigenfunction of the Fokker-Planck operator,
\beq
	\Lnoise{} \ \rho (x) = e^{-\gamma} \rho(x).
\eeq
As $\Lnoise{}$ is a non-negative operator, its leading eigenvalue is non-degenerate, real,
and positive, and the corresponding eigenvector has non-negative coordinates, by
Perron-Frobenius theorem\rf{perg,ruell68}. This property is also relevant for
the applicability of perturbation theory.

The Fokker-Planck operator and its adjoint
\beq
	\Lnoise{\dagger} \phi_n \ (y)
	= \int \phi_n (x) \ \exp\left\{-\frac{1}{2} \transp{[x-f(y)]} \diffTen^{-1}(y) [x-f(y)]\right\} \ [dx] \
\ee{FPAop}
have a whole spectrum of distinct \textit{right} and \textit{left} eigenfunctions, which contain information
on how an initial density decays to the stationary distribution,
\beq
	\Lnoise{} \ \rho_i (x) = \Gamma_i \ \rho_i(x) \hspace{2em}
	\Lnoise{\dagger} \ \tilde{\rho}_i (x) = \Gamma_i^* \ \tilde{\rho}_i (x).
\eeq

Importantly, Eq.~(\ref{FPop}) defines an integral operator with a $L_2(R^d)$
kernel (said of the Hilbert-Schmidt class\rf{Kato80}), and as such, it is
bounded and thus continuous on its support, that is $\|\rho_\epsilon~-~\rho\|~\rightarrow~0$
implies $\|\Lnoise{}\rho_\epsilon-\Lnoise{}\rho\|\rightarrow0$ and vice versa.
As mentioned in the introduction, that is the main difference with the noiseless
Perron-Frobenius operator, and the condition for us to apply perturbation theory
(details are given in~\ref{app_epsilon}).

\section{Perturbation theory}
\label{PT}
Add a perturbation to the deterministic part of the system:
\beq
	x_{n+1} = f^{(0)}(x_n) + f^{(1)}(x_n) + \xi_n.
\ee{JMH:pert_lange}
The perturbation $f^{(1)}$ is small in a sense defined later.
Without the noise, the new problem would be just as difficult as the original.
The stationary distribution of a typical chaotic system has fractal structure;
any finite perturbation results in the bifurcation of some of the infinitely many periodic orbits.
The escape rate and other observables would thus change discontinuously.
Perturbation theory would fail, in general.
Here we have noise, and
Eq.~(\ref{JMH:pert_lange}) results in a perturbation of the kernel of the original Fokker-Planck operator,
which we only keep the first order of:
\bea
	\nonumber
	\exp\left\{-\frac{1}{2}[y - f^{(0)}(x) - f^{(1)}(x)]^\top \diffTen^{-1}
		[y - f^{(0)}(x) - f^{(1)}(x)]\right\}
	 \approx  \\
		\left( 1 \ + \ [y - f^{(0)}(x)]^\top \diffTen^{-1} f^{(1)}(x) \right)
		\exp\left\{-\frac{1}{2}[y - f^{(0)}(x)]^\top \diffTen^{-1} [y - f^{(0)}(x)]\right\}.
\label{kerpert}
\eea
In order that the perturbation be small, we should have
\beq
	\left|[y - f^{(0)}(x)]^\top \diffTen^{-1} f^{(1)}(x)\right| \ll 1.
\ee{SmallDef}
Suppose that the typical scale of the dynamics in the state space is $L$,
so $|y - f^{(0)}(x)| \lesssim L$,
and the amplitude of the noise in any direction is $\langle\xi_n^{(i)}\xi_n^{(i)}\rangle= 2D$.
We are considering weak noise with $(2D)^{1/2} \ll L$.
The perturbation can be considered small if
\beq
     |f^{(1)}(x)|  \ll \frac{2D}{L}
\ee{SimpleSmallDef}
everywhere on the attractor.
We thus have three distinct length scales. The length scale
associated with the unperturbed dynamics is much longer than the
length scale associated with the noise. Both are much longer
than the length scale associated with the perturbation to the dynamics.

Now that we have a first order perturbation to the operator,
we can find the corresponding perturbations to the escape multiplier and
stationary distribution. In the light of the observations made in
the previous section on the properties of the leading eigenvalue/function of $\Lnoise{}$,
we assume in what follows
the $i^{th}$ eigenvalue
of the unperturbed Fokker-Planck operator to be simple and
non-splitting (that is, no phase transitions occur)
under perturbations. Then, we perturb the eigenvalue problem and obtain
\beq
	\big(\Lnoise{(0)} + \Lnoise{(1)}\big)
	\big(\rho^{(0)}_i + \rho^{(1)}_i \big)
	=\big(\Gamma^{(0)}_i + \Gamma^{(1)}_i\big)
	\big(\rho^{(0)}_i  + \rho^{(1)}_i \big).
\ee{Pert1}
Although this looks similar to the perturbation theory used in quantum mechanics, there are several complications.
The Fokker-Planck operator is not self-adjoint, and thus
right (left) eigenfunctions do not form an orthogonal set by themselves, but
rather, right \textit{and} left eigenfunctions form a
bi-orthogonal set, namely $\int dx \tilde{\rho}_i (x) \rho_i (x)\neq0$ only if $i=j$.

We also have a different normalization condition.
Since the leading eigenfunction is a probability distribution,
it is normalized according to $ \int \rho(x) \ dx = 1 $.
The normalization used in quantum perturbation theory is instead
$ \int \psi^*(x) \ \psi(x) \ dx = 1 $.

Take \refeq{Pert1}, keep only the first order terms,
and cancel the unperturbed eigenvalue equation.
\beq
	\Lnoise{(1)} \rho_i^{(0)} + \Lnoise{(0)} \rho_i^{(1)} =
	\Gamma_i^{(1)} \rho_i^{(0)} + \Gamma_i^{(0)} \rho_i^{(1)}
\ee{FirstOrderTerms}
To find the first order correction to the eigenvalue,
multiply by the corresponding unperturbed left eigenfunction and
integrate over the entire space.
\[
	\int \tilde{\rho}_i^{(0)} (x)^* \Lnoise{(1)} \rho_i^{(0)} (x) \ dx \ + \
	\int \tilde{\rho}_i^{(0)} (x)^* \Lnoise{(0)} \rho_i^{(1)} (x) \ dx \ = \
	\Gamma_i^{(1)} \int \tilde{\rho}_i^{(0)} (x)^* \rho_i^{(0)} (x) \ dx \ + \
	\Gamma_i^{(0)} \int \tilde{\rho}_i^{(0)} (x)^* \rho_i^{(1)} (x) \ dx 
\]
\beq
	\int \tilde{\rho}_i^{(0)} \Lnoise{(1)} \rho_i^{(0)}  +
	\int \tilde{\rho}_i^{(0)} \Lnoise{(0)} \rho_i^{(1)}  =
	\Gamma_i^{(1)} \int \tilde{\rho}_i^{(0)}  \rho_i^{(0)}  +
	\Gamma_i^{(0)} \int \tilde{\rho}_i^{(0)} \rho_i^{(1)}
\eeq
\beq
\Longrightarrow	\int \tilde{\rho}_i^{(0)} \Lnoise{(1)} \rho_i^{(0)}  =
	\Gamma_i^{(1)} \ \int\tilde{\rho}_i^{(0)}  \rho_i^{(0)} .
\eeq
The first order correction to the eigenvalue is thus:
\beq
	\Gamma_i^{(1)} = \frac{\int \tilde{\rho}_i^{(0)} \Lnoise{(1)} \rho_i^{(0)}}{\int \tilde{\rho}_i^{(0)} \rho_i^{(0)}}.
\ee{EigVal}
We recall that the theory presented so far applies to a non-negative operator of the
Hilbert-Schmidt class, thus infinite dimensional.
In order to find the first order correction to the eigenfunction, rewrite
\refeq{FirstOrderTerms} so that each side looks like an eigenvalue equation.
\beq
	\left(\Lnoise{(0)} - \Gamma_i^{(0)} \matId \right) \rho_i^{(1)} =
	-\left(\Lnoise{(1)} - \Gamma_i^{(1)} \matId \right) \rho_i^{(0)}
\eeq
To isolate $\rho_i^{(1)}$, we would like to invert
$\left(\Lnoise{(0)} - \Gamma_i^{(0)} \matId \right)$.
Unfortunately, this is guaranteed to not be invertible since $\Gamma_i^{(0)}$
is an eigenvalue of $\Lnoise{(0)}$.
If the Fokker-Planck operator were self-adjoint like in quantum mechanics,
the eigenfunctions would form a complete basis, and
the perturbation to the eigenfunction
could be written
as a superposition of the unperturbed eigenfunctions, except the one that we
are calculating the perturbation to.
The orthogonality of the eigenfunctions would then allow us to isolate the coefficients
of the first order correction of the eigenfunction in this basis of unperturbed
eigenfunctions (see for example\rf{sakurai}).
But since $\Lnoise{(0)}$ is neither Hermitian nor in general diagonalizable,
we may not proceed in this fashion, and instead
we get around the problem by using the Moore-Penrose
pseudoinverse\rf{BarHuss11},
$\left(\Lnoise{(0)} - \Gamma_i^{(0)} \matId\right)^+$,
that first projects the subspace
of $\Gamma_i^{(0)}$  out of $\Lnoise{(0)}~-~\Gamma_i^{(0)}~\matId$,
and then takes the inverse of the resulting operator. In order
to evaluate this pseudoinverse, we have to introduce some truncated basis for $L_2(R^d)$,
and thus restrict our analysis henceforth to a finite-dimensional subspace.
The operator equation \refeq{FINAL} is then reduced to a matrix equation which can
be solved numerically.
The details of this procedure are shown in~\ref{app_matrices}.
At the end, we obtain
\beq
	\rho_i^{(1)} = -\left(\Lnoise{(0)} - \Gamma_i^{(0)} \matId \right)^+
	\Lnoise{(1)} \rho_i^{(0)} .
\ee{FINAL}
Note that the normalization of the unperturbed eigenfunction is irrelevant.
Multiplying the unperturbed eigenfunction by a constant simply multiplies the
perturbation by the same constant, so the relative size of the perturbation
remains unchanged.

The properties of the Moore-Penrose operator and its relation
to other pseudoinverses are reported in~\ref{MPvsOth}.

\section{Numerical results for the Lozi map}
\label{numcs}
We test this technique on the Lozi map with a deterministic correction $q(x,y)$,
and weak, white noise:
\bea
	\nonumber
	x_{n+1} &=& 1 - a \ |x_n| + b \ y_n + \epsilon \ q(x_n,y_n) +\xi_n^{(x)} \\
	y_{n+1} &=& x_n + \xi_n^{(y)}
	\label{lozi_map}
\eea
The random variable added to this system is uniformly distributed according to
\beq
	P(\xi) = \frac{1}{ \sqrt{\det{(2 \pi \Delta)}}}
	\exp{\Big(-\frac{1}{2} \transp{\xi} \diffTen^{-1} \xi \Big)} .
\eeq
Here $\diffTen$ is isotropic. Its magnitude varies over the course of this study.

\begin{figure}
	\centering
	(a)\includegraphics[width = .35\textwidth]{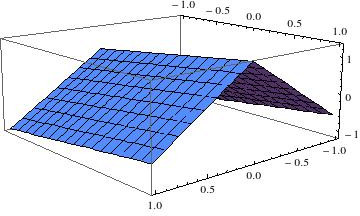}
	(b)\includegraphics[width = .45\textwidth]{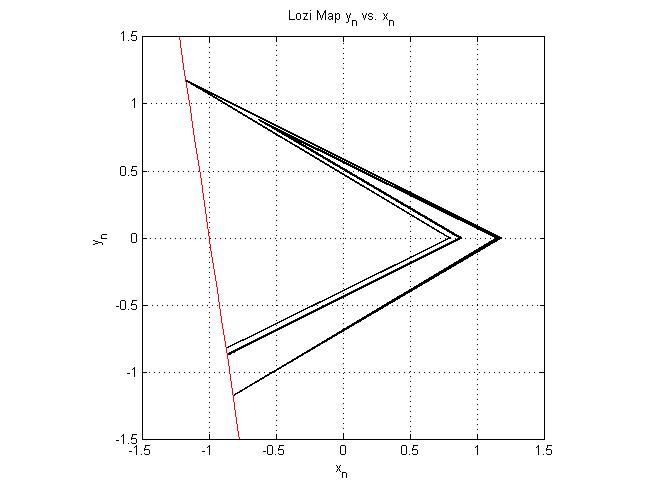}
	\caption{(a) $f(x,y)=1-a|x|+by$ shows the `tent' shape of the Lozi map,
		     its slope controlled by the parameter $a$; (b)
		    the Lozi attractor with parameters $a = 1.85, b = 0.3$
	     calculated using the long time behavior of a single generic orbit.
	     The red line shows the boundary of the basin of attraction.
	   Any points to the left of that line escape to $(-\infty,-\infty)$
	 and any points to the right approach the attractor
	 as the map is iterated without noise.}
	\label{lozimap}
\end{figure}

\begin{figure}
        \centering{
        (a)\includegraphics[width = .25\textwidth]{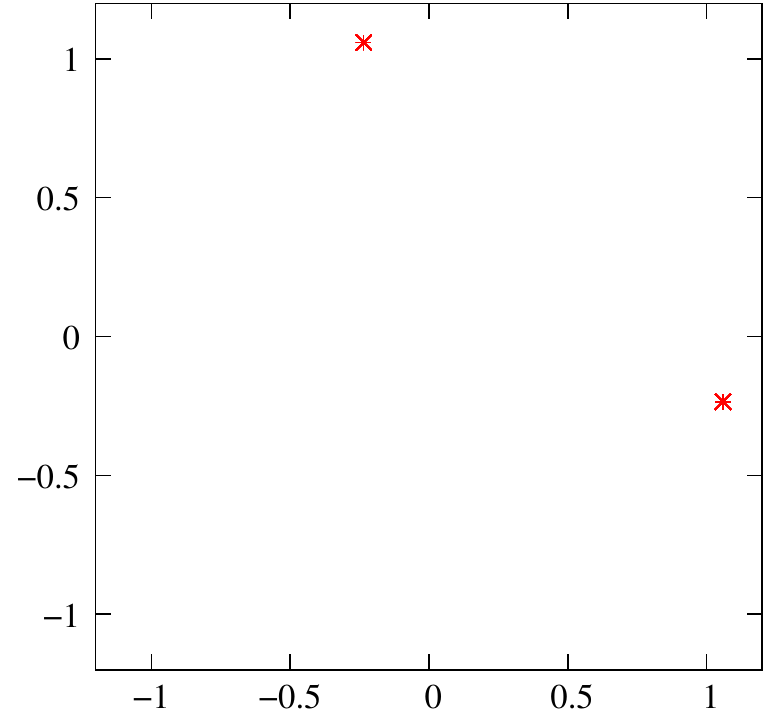}
        (b)\includegraphics[width = .25\textwidth]{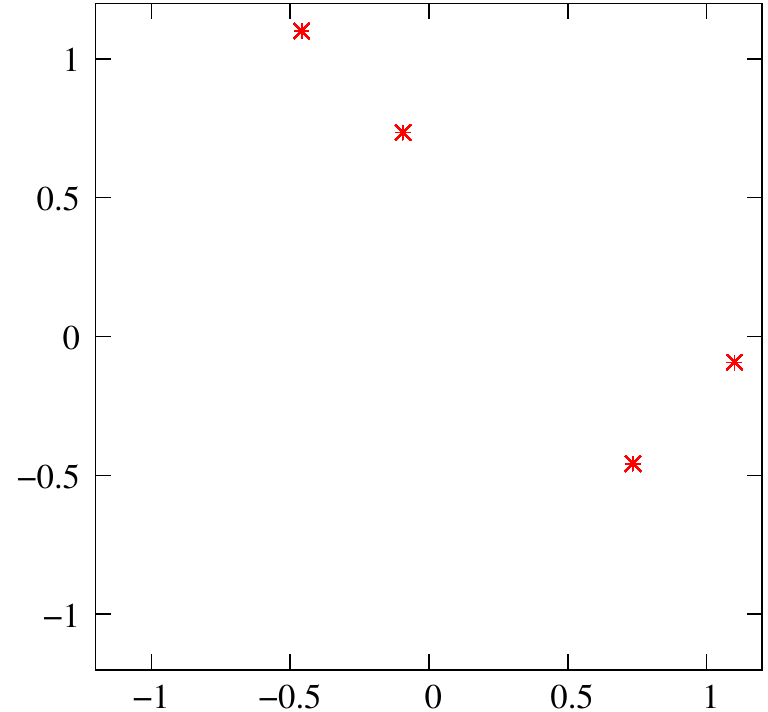}
	(c)\includegraphics[width = .25\textwidth]{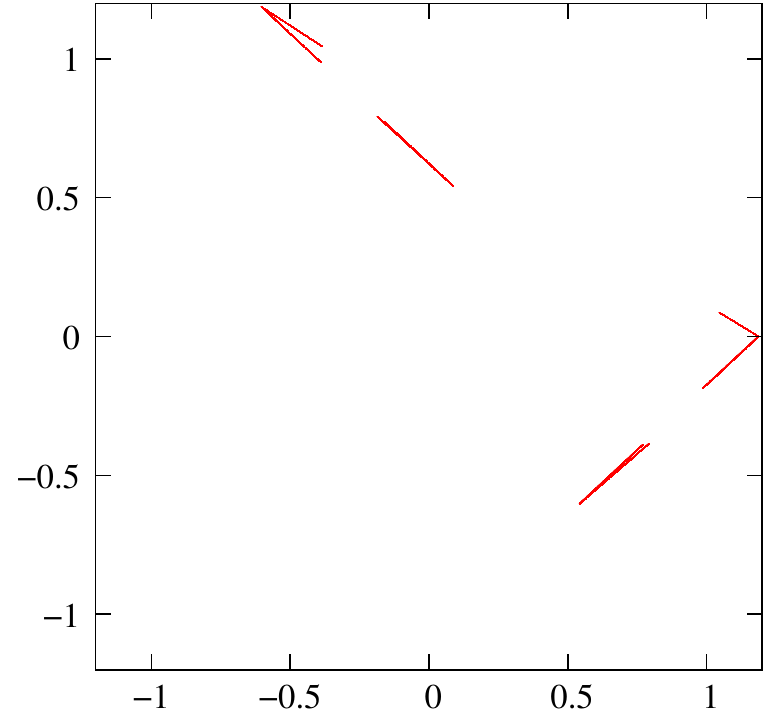}}
	\centering{
	(d)\includegraphics[width = .25\textwidth]{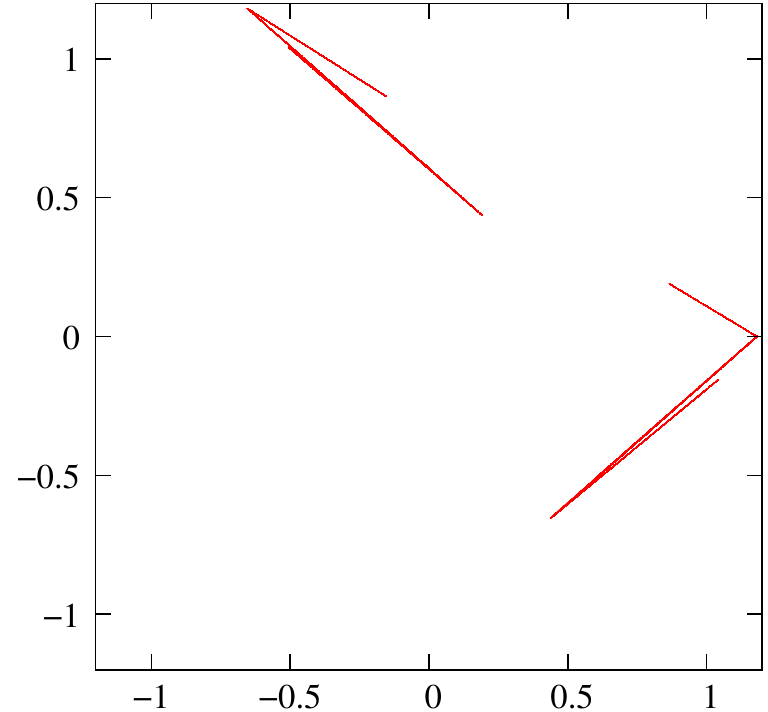}
	(e)\includegraphics[width = .25\textwidth]{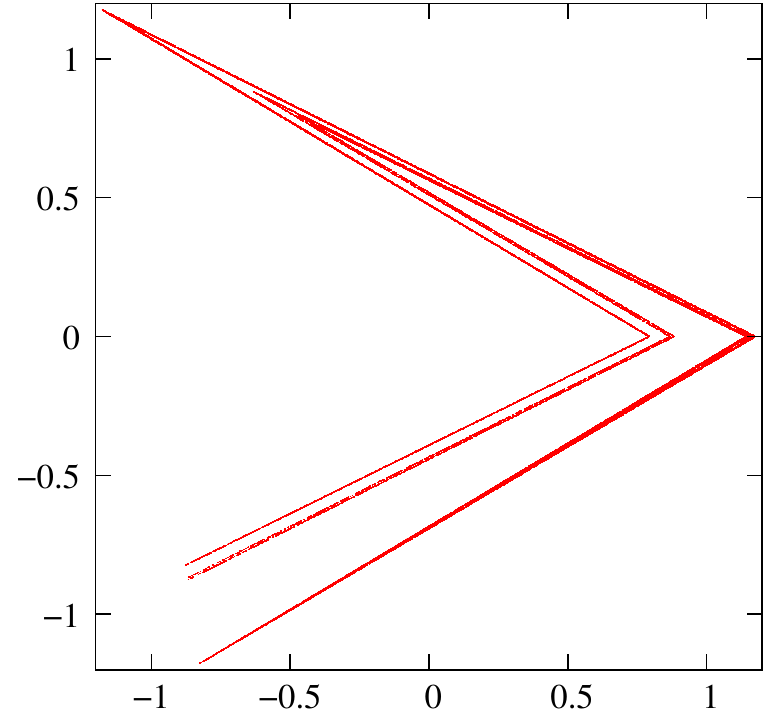}
	(f)\includegraphics[width = .25\textwidth]{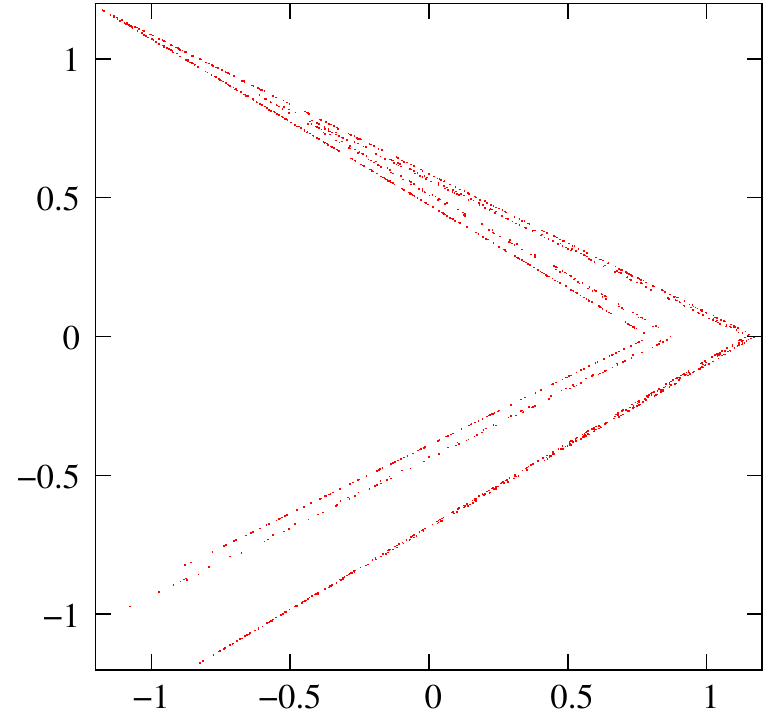}}
        \caption{Snapshots of the bifurcation cascade of the Lozi map,
		from (a) a period-2 stable cycle for $a=1$, to:
		(b) a period-4 stable cycle ($a=1.3$); (c) a
		four-region chaotic attractor ($a=1.35$);
		(d) a two-region chaotic attractor ($a=1.4$);
		(e) the one-region attractor used in this study ($a=1.85$),
		that quickly fades into a repeller when the parameter is
		further increased, as in (f) ($a=1.855$).}
        \label{lozicasc}
\end{figure}

The Lozi map is a two-dimensional version of the tent map [\reffig{lozimap}(a)],
and it undergoes a bifurcation cascade as the slope
(controlled by the parameter $a$) is varied
(\reffig{lozicasc}). Here we set
$a = 1.85$ and $b = 0.3$, corresponding to chaotic dynamics
confined in the attractor of \reffig{lozimap}(b).
The additive perturbation $q(x,y)$ is a smooth function, while
$\epsilon$ is the small parameter and can be either positive or negative.
In what follows, we will take $q(x,y)=x^2$ (analyses of the
cases $q(x,y)=x^3$ and $q(x,y)=y^2$
can be found in~\ref{otherpert}), 
which is of particular interest, being qualitatively
equivalent to a perturbation of the parameter $a$, where
the bifurcations are understood. So, for example, when $\epsilon>0$,
adding $\epsilon x^2$ to the Lozi map is equivalent to
decreasing the parameter $a$ and thus slightly
deforming the attractor of~\reffig{lozimap}(b) [or~\reffig{lozicasc}(e)],
whereas $\epsilon<0$ effectively
increases the parameter $a$, so as to turn the Lozi attractor
of~\reffig{lozicasc}(e) into a repeller [\reffig{lozicasc}(f)].

We divide the region $\pS=[-1.5,1.5] \times [-1.5,1.5]$ into 3600 uniform grid elements,
and we call each $\pS_k$.
When calculating the perturbation to the eigenfunction,
we use the basis of characteristic functions for each of our grid elements.
\beq
	\chi_k (x) \equiv \left\{
		\begin{array}{cl}
			\frac{1}{\sqrt{|\pS_k|}} & \mbox{if $x\in\pS_k$} \\
			0 & \mbox{otherwise .}
		\end{array}
	\right.
\eeq

We want to compute the leading
eigenvalue and eigenfunction from a numerical approximation of the one-step transfer matrix\rf{Ulam60},
that is the discretized Fokker-Planck operator, whose entries are
probabilities of a trajectory to hop between any pair of grid
elements in one time step.
The transfer matrix is given by 
\beq
        \Lnoise{}_{ij}=
	\int_{\pS_i}[dy]
	\int_{\pS_j}[dx]
	\exp\left\{-\frac{1}{2} \transp{[y-f(x)]} \diffTen^{-1}(x) [y-f(x)]\right\}
\label{Lmat}
\eeq
that in general would be a $2N$-dimensional integral (4D in our model). Since the
kernel of the operator is smooth, and
it varies on a length scale longer than the grid mesh, as long as
the amplitude of the noise in our simulations is
large enough, we may approximate~\refeq{Lmat} with
\beq
 	\Lnoise{}_{ij}\simeq \exp\left\{-\frac{1}{2} \transp{[y_j-f(x_i)]} \diffTen^{-1} [y_j-f(x_i)]\right\},
\label{app_Lmat}
\eeq
where $x_i$ and $y_j$ are the centers of the corresponding grid elements\footnote{
As the grid is uniform, we can set the area of the mesh to unity in~\refeq{app_Lmat}
and then just normalize the leading eigenvector of $\Lnoise{}_{ij}$.
}.
We may ensure the validity of the above approximation by
estimating the error from the first-order Taylor expansion of the integrand
in~\refeq{Lmat}, and applying the mean value theorem for integrals, so that
the ratio of
the overall remainder of the integral
to the leading term~\refeq{app_Lmat}
is upper bounded by
\beq
|l^2|\sup_{\pS_i,\pS_j}\left|\nabla^2\exp\left\{-\frac{1}{2} \transp{[y-f(x)]}\diffTen^{-1}(x) [y-f(x)]\right\}\right|.
\ee{int_rem}
The size $l$ of the mesh in the discretization is $O(10^{-2})$, while $f(x)$ is (piecewise) linear and
at most $O(1)$ on the Lozi attractor, hence~\refeq{int_rem} is $O(10^{-4})$ or less, orders of magnitude
smaller than the typical size of the perturbation in Eq.~\refeq{kerpert}.

The Lozi map is not globally attracting.
The stable manifold of one of the two fixed points acts as
a boundary for the basin of attraction (see~\reffig{lozimap}).
Without the noise, any orbit which begins in the basin of attraction will
stay on the attractor for all time.
The noise allows orbits to cross this boundary and results in a nonzero escape rate.
Once an orbit crosses the deterministic boundary,
it accelerates off to $(-\infty,-\infty)$.
To account for this, we say that an initial condition escapes in one time step
if it crosses the deterministic boundary of the basin of attraction.
These points are excluded from the calculation of the transfer matrix~\refeq{app_Lmat}.
Once we have the transfer matrix, we calculate the eigenvalues and eigenvectors.
The largest magnitude eigenvalue is the escape multiplier $\Gamma_0 = e^{-\gamma}$,
where $\gamma$ is the escape rate.
Since our system is ergodic, the largest magnitude eigenvalue is
guaranteed to be real and isolated\rf{ruelle}.

Our goal is to estimate the escape rate, as well as the stationary distribution
of~\refeq{lozi_map}
to the first non-trivial order of perturbation, according to~\refeq{EigVal}
and~\refeq{FINAL}, respectively. We utilize in these formulae the stationary
distributions $\rho^{(0)}(x)$ and $\tilde{\rho}^{(0)}(x)$ of the unperturbed ($\epsilon=0$)
Fokker-Planck operator
$\Lnoise{(0)}$
and its adjoint $\left[\Lnoise{(0)}\right]^\dagger$, respectively, computed by means of~\refeq{app_Lmat}.
On the other hand, we apply the same discretization to diagonalize the
transfer matrix of the perturbed map~\refeq{lozi_map} ($\epsilon\neq0$), as a way to
check the accuracy of the
perturbation theory calculation.

\begin{figure}
	\centering{
	(a) \includegraphics[width = .45\textwidth]{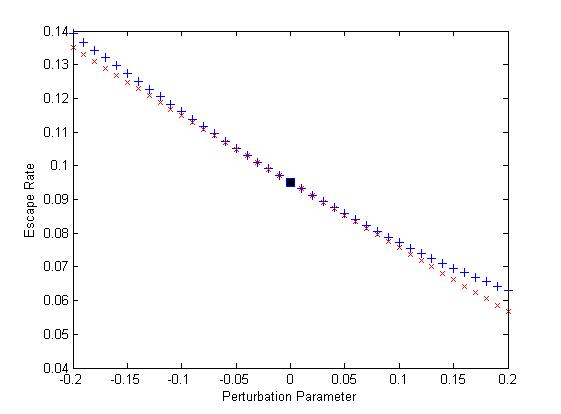}
	(b)	\includegraphics[width = .45\textwidth]{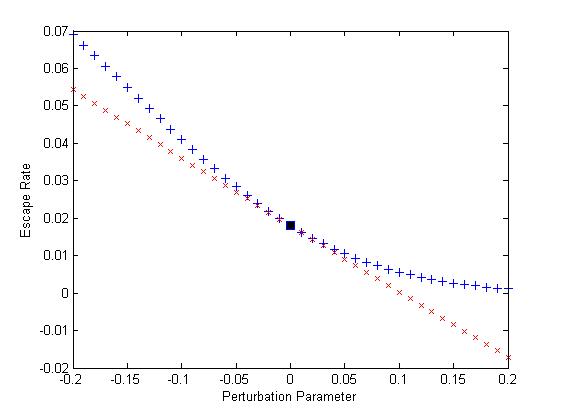}}
	\caption{The escape rate $\gamma$ from the Lozi attractor, as a function of
	the perturbation parameter $\epsilon$, for:
	(a) $\diffTen = .1 \ \matId$ and (b) $\diffTen = .01 \ \matId$.
        Blue pluses: exact calculation.
	Red crosses: perturbation theory.}
	\label{ERvsEps}
\end{figure}

We perform our calculations for two different noise amplitudes, $\diffTen = 0.1 \ \matId$ and
$\diffTen = 0.01 \ \matId$, to investigate how the strength of the noise
influences the efficacy of perturbation theory, and we
compute the escape rate for a range of values of the small parameter $\epsilon$,
controlling the size of the deterministic perturbation.
Based on
Eq.~(\ref{SimpleSmallDef}),
we can predict roughly that the perturbation theory will break down
at $\epsilon\sim2D/L\approx 0.05,0.005$, depending on the
noise amplitude.
\refFig{ERvsEps} illustrates the results:
the escape rate computed with
perturbation theory [\refeq{EigVal}] is in agreement with the result from the diagonalization
of the transfer matrix~\refeq{Lmat} for the full map~\refeq{lozi_map},
for a range of $\epsilon$ about an order of magnitude larger than
what estimated with Eq.~(\ref{SimpleSmallDef}).

\begin{figure}
	\centering
	(a)\includegraphics[width=.45\textwidth]{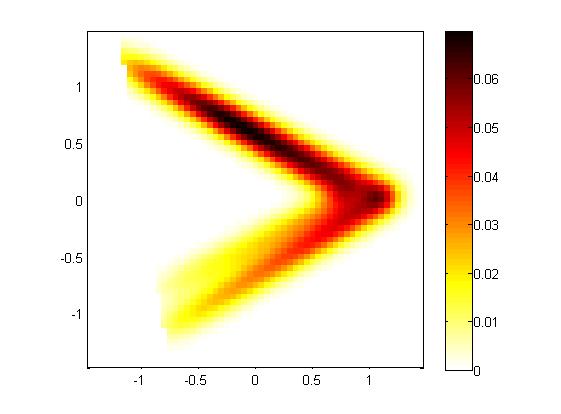} \\
	(b)\includegraphics[width=.45\textwidth]{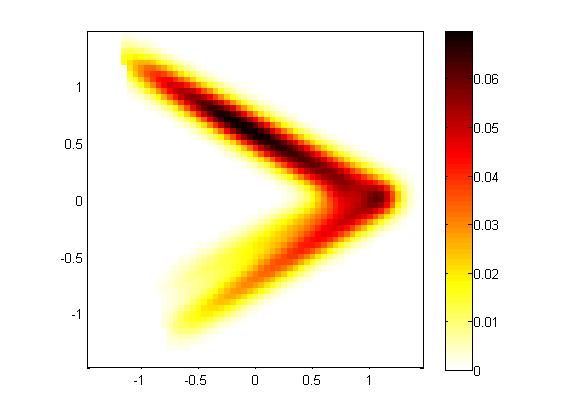}
	(c)\includegraphics[width=.45\textwidth]{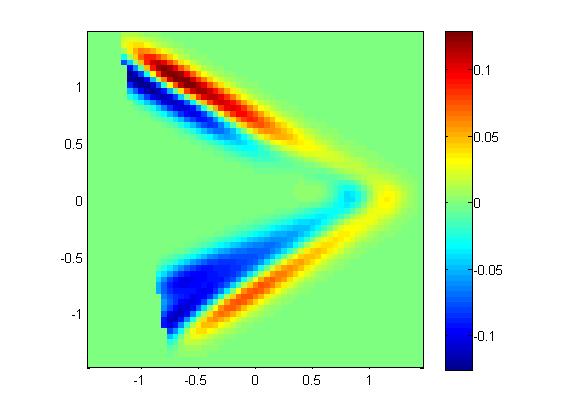} \\
	(d)\includegraphics[width=.45\textwidth]{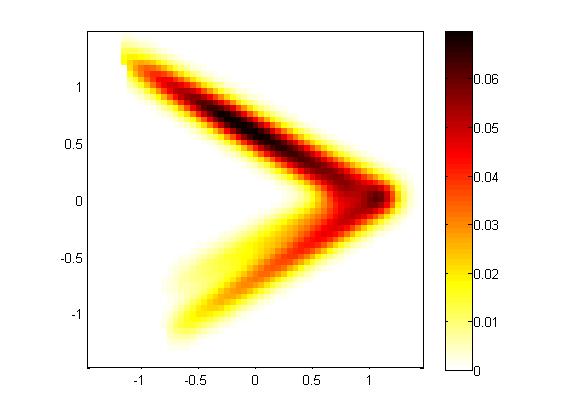}
	(e)\includegraphics[width=.45\textwidth]{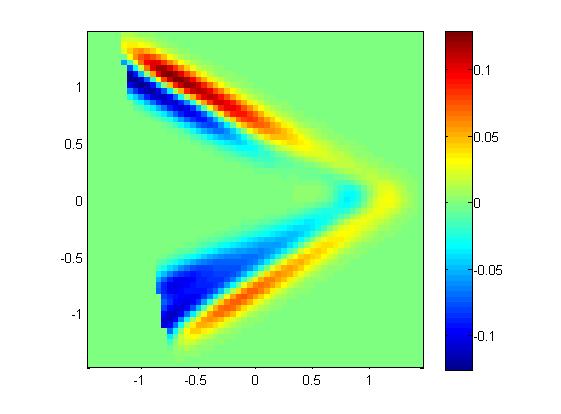} \\
	\caption{Density plots of the stationary distributions for $\diffTen = .01 \ \matId$
	and $\epsilon = .05$.
	(a) $\msr^{(0)}(x)$.
	(b) $\msr(x)$.
	(c) $\delta\msr(x)$.
	(d) $\msr^{(1)}(x)$.
	(e) $\delta\msr^{(1)}(x)$.}
	\label{PTCompareD010}
\end{figure}

\begin{figure}
	\centering
	(a)\includegraphics[width=.45\textwidth]{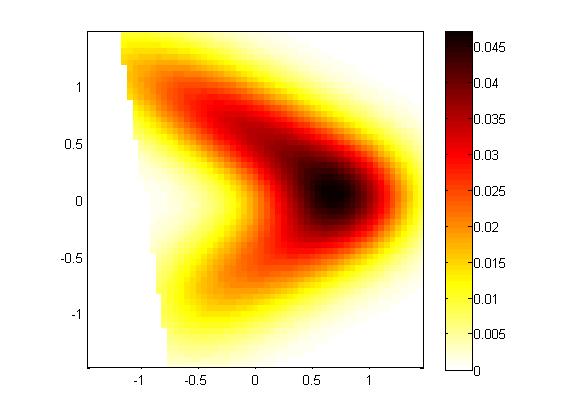} \\
	(b)\includegraphics[width=.45\textwidth]{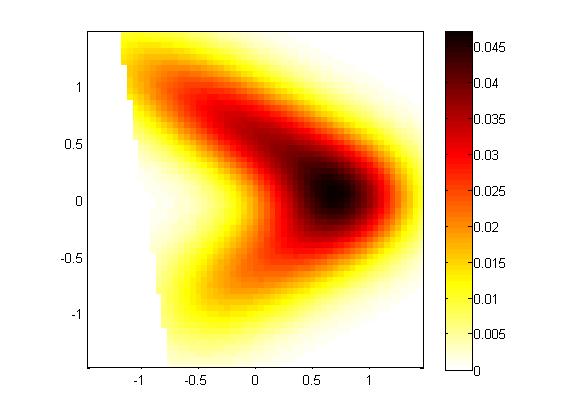}
	(c)\includegraphics[width=.45\textwidth]{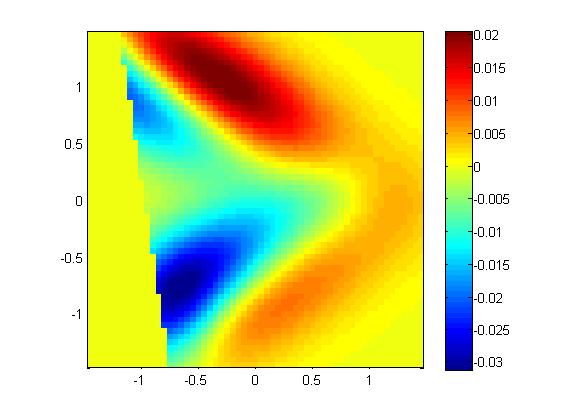} \\
	(d)\includegraphics[width=.45\textwidth]{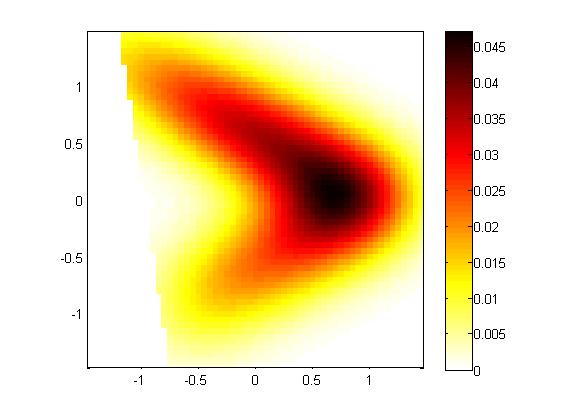}
	(e)\includegraphics[width=.45\textwidth]{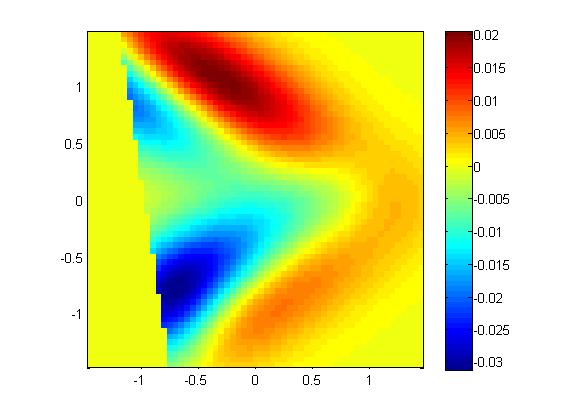} \\
       \caption{Density plots of the stationary distributions for $\diffTen = .1 \ \matId$
        and $\epsilon = .05$.
        (a) $\msr^{(0)}(x)$.
        (b) $\msr(x)$.
        (c) $\delta\msr(x)$.
        (d) $\msr^{(1)}(x)$.
        (e) $\delta\msr^{(1)}(x)$.}
	\label{PTCompareD100}
\end{figure}

\begin{figure}
	\centering{
	(a) \includegraphics[width = .45\textwidth]{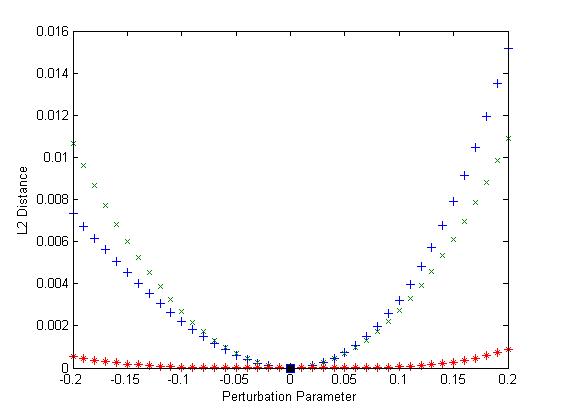}
	(b)	\includegraphics[width = .45\textwidth]{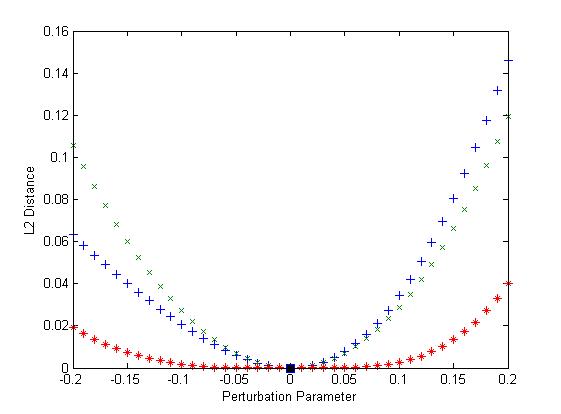}}
	\caption{$L_2$ distance $d$ between various distributions as a function of the
	perturbation parameter $\epsilon$.
	Blue pluses: $d(\msr^{(0)},\msr)$.
	Green crosses: $d(\msr^{(0)},\msr^{(1)})$.
	Red asterisks: $d(\msr^{(1)},\msr)$.
	(a) $\diffTen = .1 \ \matId$ and (b) $\diffTen = .01 \ \matId$.}
	\label{L2vsEps}
\end{figure}

Next, the stationary distribution $\rho^{(1)}(x)$ is computed to the first non-trivial
perturbative order using~\refeq{FINAL}, and then again compared with
the first eigenfunction $\rho(x)$ of the spectrum of~\refeq{app_Lmat}, by means of the
$L_2$ distance
\beq
 d(\msr,\msr^{(1)}) =       \frac{\int \left[ \msr(\ssp) - \msr^{(1)}(\ssp) \right]^2 dx}
        {\int \left[ \msr(\ssp) \right]^2 d\ssp}.
\ee{L2Diff}
Density plots of the calculated distributions 
are shown
in \reffig{PTCompareD010} and \reffig{PTCompareD100}.
Due to the small value of $\epsilon$, the
stationary distributions $\msr(\ssp), \msr^{(1)}(\ssp)$ of the perturbed map
can hardly be distinguished from
the invariant density $\msr^{(0)}(\ssp)$ of the unperturbed system.
To make the distinctions more apparent, we also plot
the differences
\bea
\delta\msr(\ssp) &=& \frac{\msr(\ssp)-\msr^{(0)}(\ssp)}{\epsilon}, \\
\delta\msr^{(1)}(\ssp) &=& \frac{\msr^{(1)}(\ssp)-\msr^{(0)}(\ssp)}{\epsilon}.
\label{msr_diffs}
\eea
As shown in the figures,
the differences $\delta\msr(\ssp)$ and $\delta\msr^{(1)}(\ssp)$
look almost identical, as evidence that $\msr^{(1)}(\ssp)$ is
a good estimate for $\msr(\ssp)$.
The range of validity of the perturbative approximation for the stationary
distribution is probed in~\reffig{L2vsEps}, where
the $L_2$ distances $d(\msr^{(0)},\msr)$, $d(\msr^{(0)},\msr^{(1)})$,
and $d(\msr^{(1)},\msr)$  are plotted
as a function  of $\epsilon$.
It is apparent from the graph that
$d(\msr^{(0)},\msr)\sim\epsilon^2$, and $d(\msr^{(0)},\msr^{(1)})\sim\epsilon^2$,
which is in agreement with the expectation
$\|\Lnoise{}_\epsilon\rho-\Lnoise{}\rho\|\sim\epsilon$
(proven in~\ref{app_epsilon}).
Surprisingly, instead, the $L_2$ distance
$d(\msr^{(0)},\msr^{(1)})\ll\epsilon$
for a range of $|\epsilon|$ that largely exceeds the expectations.

\section{Conclusions, comments}
\label{summ}

The fractal structure of the stationary distribution
for a system exhibiting deterministic chaos changes discontinuously in
response to any finite perturbation, in principle
preventing any perturbative calculations.

In contrast, weak noise smears out these fractals.
This is apparent in a Fokker-Planck picture: the evolution
operator becomes continuous,
the stationary distribution (natural measure)  is smooth,
and, consequently, the system lends itself to a perturbative treatment.

We have expanded the Fokker-Planck evolution operator of a noisy map
subject to a deterministic perturbation 
in a power series.
The long-time observables of the perturbed
system can be estimated in terms of the unperturbed one.
As an example, we have obtained excellent approximations
to the escape rate and to the stationary distribution of
a noisy Lozi attractor, valid for a range of the
perturbation parameter proportional to the amplitude of the
noise, and always at least one order of magnitude larger
than expected.

The successful implemetation  of perturbation
theory in a chaotic system affected by weak noise
can also be relevant for the following reasons:
\begin{enumerate}
\item the relative insensitivity to fluctuations demonstrated
constitutes evidence
that a chaotic system acquires a \textit{finite} resolution
when noise is introduced,
and it can be modelled via a transfer operator of
finite degrees of freedom. This problem is discussed
extensively in\rf{LipCvi08,CviLip12,HenCviLip14};
\item the analysis presented sets bounds
for the robustness of a model, and it is straightforward enough to
be implemented in any algorithm that creates a
template out of a time series. It is noted that
low-dimensional, noisy discrete-time mappings are
still widely used as models
in several fields of science and
engineering\rf{WeezCoupLas,BeiramiZigzag,CardBifc}.
\end{enumerate}
Future work in this direction may investigate
\begin{enumerate}
\item the validity of
higher-order  expansions, which we do not expect to change
the results reported here qualitatively, at least for the
leading eigenvalue/function of the Fokker-Planck operator.
The reason is  that  perturbation series are trivial about simple, non-splitting
eigenvalues, while one has fractional (`Puiseux'\rf{Kato80}) power
series in the vicinity of phase transitions, with successive terms (much)
less than an order of magnitude apart from one another;
\item evolution operators which carry finite memory of  the past
(for example of the Mori-Zwanzig type~\rf{Chorin02,VentKar14}),
as opposed to the Markovian operators considered here, and whether and
to what extent
those lend themselves to a perturbative approach.
Markovian operators other than the Fokker-Planck operator are implemented
in the same way: just replace the Gaussian in \refeq{FPop} and \refeq{FPAop} 
with a different kernel.
\end{enumerate}

\section{Acknowledgments}
D.L. acknowledges the National Science Fundation of
China (NSFC) for partial support (Grant No. 11450110057-041323001).
J.M.H. thanks D.L. and Prof. Nianle Wu for the hospitality at IASTU, Beijing,
where part of this work was done.
P.C. thanks the family of late G. Robinson, Jr. and NSF grant DMS-1211827
for partial support.

\appendix
\section{Additional perturbations}
\label{otherpert}
\begin{figure}
        \centering
        (a) \includegraphics[width=.45\textwidth]{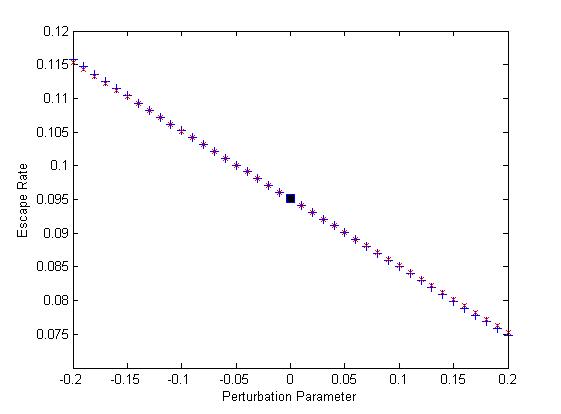}
        (b) \includegraphics[width=.45\textwidth]{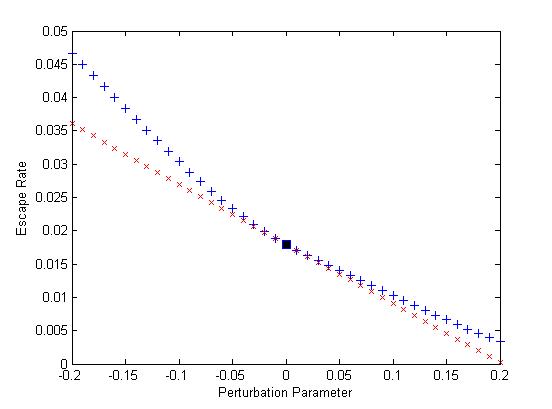} \\
        (c) \includegraphics[width=.45\textwidth]{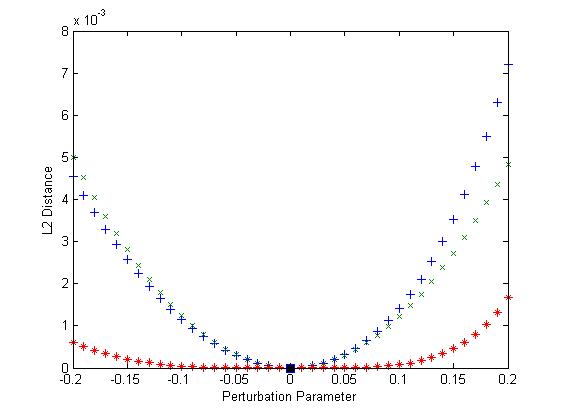}
        (d) \includegraphics[width=.45\textwidth]{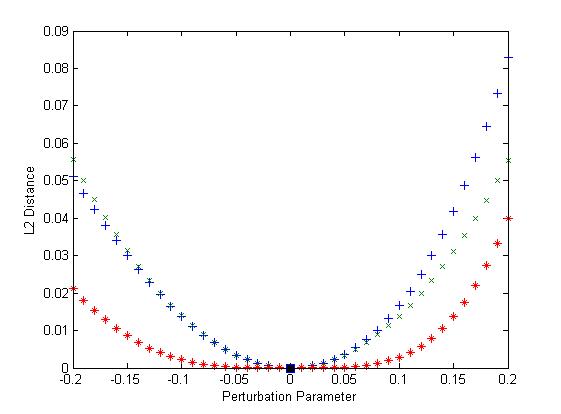}
        \caption{[(a),(b)] Escape rate $\gamma$ vs. perturbation parameter $\epsilon$
        for a perturbation of the form $q(x,y)=x^3$. Blue pluses: exact calculation.
	Red crosses: perturbation theory. (a) $\diffTen = .1 \matId$ (b) $\diffTen = .01 \matId$.
        [(c),(d)] $L^2$ distance $d$ vs. $\epsilon$.
	 Blue pluses: $d(\msr^{(0)},\msr)$.
        Green crosses: $d(\msr^{(0)},\msr^{(1)})$.
        Red asterisks: $d(\msr^{(1)},\msr)$.
        (c) $\diffTen = .1 \matId$ (d) $\diffTen = .01 \matId$.}
        \label{XCubed}
\end{figure}

\begin{figure}
        \centering
        (a) \includegraphics[width=.45\textwidth]{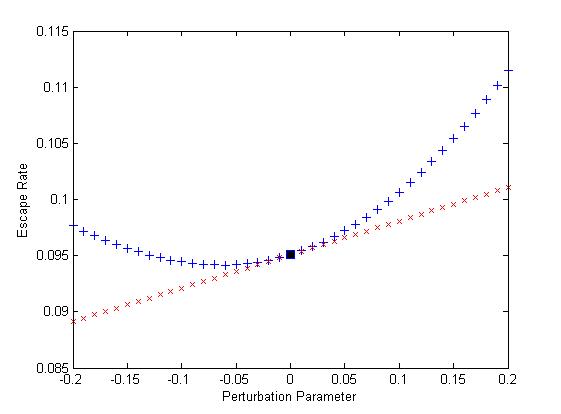}
        (b) \includegraphics[width=.45\textwidth]{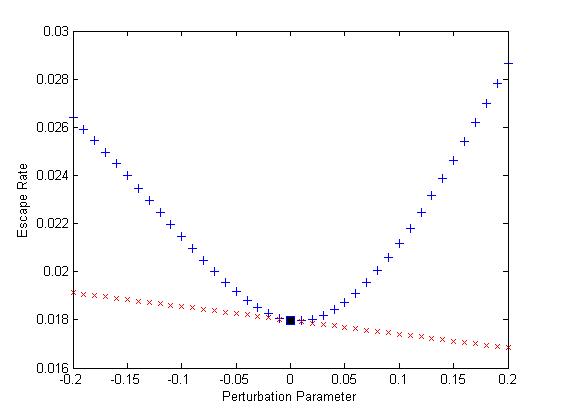} \\
        (c) \includegraphics[width=.45\textwidth]{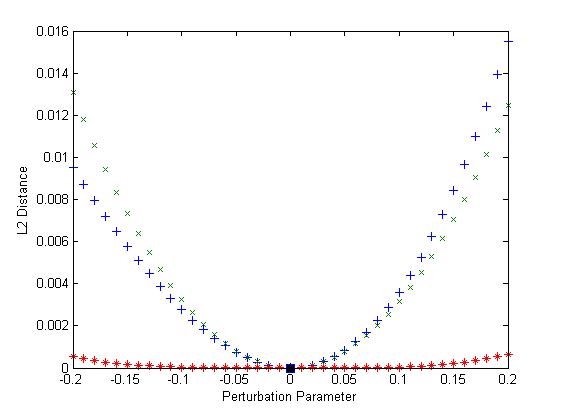}
        (d) \includegraphics[width=.45\textwidth]{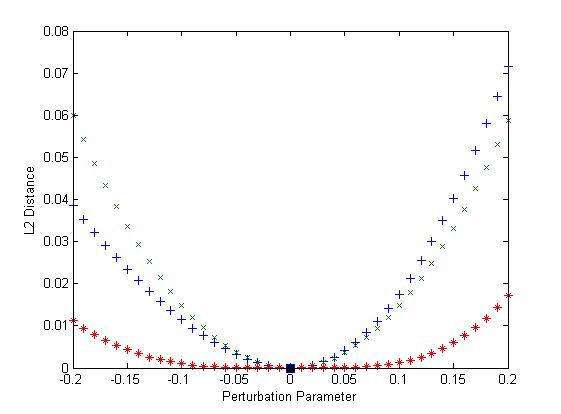}
\caption{[(a),(b)] Escape rate $\gamma$ vs. perturbation parameter $\epsilon$
        for a perturbation of the form $q(x,y)=y^2$. Blue pluses: exact calculation.
        Red crosses: perturbation theory. (a) $\diffTen = .1 \matId$ (b) $\diffTen = .01 \matId$.
        [(c),(d)] $L^2$ distance $d$ vs. $\epsilon$.
         Blue pluses: $d(\msr^{(0)},\msr)$.
        Green crosses: $d(\msr^{(0)},\msr^{(1)})$.
        Red asterisks: $d(\msr^{(1)},\msr)$.
        (c) $\diffTen = .1 \matId$ (d) $\diffTen = .01 \matId$.}
        \label{YSquared}
\end{figure}

We present in this section the outcomes of additional
tests of perturbation theory, with corrections of the forms
$q(x,y)=x^3$ (\reffig{XCubed}) and $q(x,y)=y^2$ (\reffig{YSquared})
for the Lozi map.

\section{Continuity of the Fokker-Planck operator and response to perturbations}
\label{app_epsilon}

The Fokker-Planck operator $\Lnoise{}$, supported on a bounded interval $I$,
defines a bounded mapping of $L_2(I)$ into itself. In fact, calling the kernel
\beq
L(x,y) = \frac{1}{|\det(2 \pi\diffTen(x))|^{1/2}} \exp\left\{-\frac{1}{2} \transp{(y-f(x))} \diffTen^{-1}(x) (y-f(x))\right\},
\ee{FPker}
we may write
\bea
\nonumber
\|\Lnoise{}\rho\|^2 &=& \int\left|\int L(x,y)\rho(y)dy\right|^2dx
\\
&\leq& \int\left(\int|L(x,y)|^2dy\right)\left(\int|\rho(y)|^2dy\right)dx
=  \|L\|^2\|\rho\|^2 < \infty,
\label{FPbnd}
\eea
where $\rho$ is supposed $L_2$, and all integrals are taken on a compact interval.
We want to show that a perturbation of $O(\epsilon)$ produces a response of the same order.
That is
\beq
\|\Lnoise{}_\epsilon\rho-\Lnoise{}\rho\|\sim\epsilon,
\ee{FPresp}
where
\beq
\Lnoise{}_\epsilon\rho =
 \int \rho(x) \ \exp\left\{-\frac{1}{2}\transp{(y-f(x)-\epsilon f^1(x))}\diffTen^{-1}(x) (y-f(x)-\epsilon f^1(x))\right\} \ [dx].
\ee{FPeps}
The exponential form allows us to separate the kernel and recast the $\epsilon-$dependence into the distribution it acts on
\beq
\Lnoise{}_\epsilon\rho = \Lnoise{}\rho_\epsilon,
\ee{eps_cast}
and we may rewrite
\bea
\nonumber
\|\Lnoise{}_\epsilon\rho-\Lnoise{}\rho\|&=&\|\Lnoise{}\rho_\epsilon-\Lnoise{}\rho\|
\leq \|L\| \, \|\rho_\epsilon-\rho\|  \\
&\simeq& \|L\|  \, \|[1-\epsilon yf^1(x)]\rho-\rho\| = \epsilon \|L\| \, \|yf^1(x)\rho\|,
\label{Oeps}
\eea
that is of $O(\epsilon)$, as required.

\section{First order corrections in terms of a particular basis}
\label{app_matrices}

Choose a basis to represent everything in: $\{H_k\}$, where
\beq
	\int H_j \ H_k \ dx = \delta_{jk} \nonumber .
\eeq
Note that this is not the same normalization that is used to make the
leading eigenvector a probability distribution.
This means that the coefficients of the distributions in this representation have
units, but leads to no other negative side effects.

Write both the perturbed and unperturbed eigenfunctions in this basis.
\bea
	\rho_i^{(0)}(x) &=& \sum_k P^{(0)}_{ik} \ H_k(x) \\
	\rho_i^{(1)}(x) &=& \sum_k P^{(1)}_{ik} \ H_k(x)
\eea
We will also need to have the unperturbed adjoint eigenfunction
using the same basis.
\beq
	\tilde{\rho}_i^{(0)}(x) = \sum_k \tilde{P}^{(0)}_{ik} \ H_k(x)
\eeq

The unperturbed Fokker-Planck operator
or perturbation to the Fokker-Planck can be written using a kernel.
\beq
	\Lnoise{} \ \rho_i (y) = \int L (y,x) \ \rho_i (x) \ dx
\eeq
These kernels can also be written in terms of our basis.
\bea
	L^{(0)} (y,x) &=& \sum_{jk} L^{(0)}_{jk} \ H_j(y) \ H_k(x) \\
	L^{(1)} (y,x) &=& \sum_{jk} L^{(1)}_{jk} \ H_j(y) \ H_k(x)
\eea
The action of an operator acting on an eigenfunction in terms of the basis is:
\bea
	\Lnoise{}\  \rho_i (y) &=& \int L (y,x) \ \rho_i(x) \ dx \\
	&=& \int \left( \sum_{jk} L_{jk} \ H_j(y) \ H_k(x) \right) \
			 \left( \sum_l P_{il} \ H_l(x) \right) \ dx \\
	&=& \sum_{jkl} P_{il} \ L_{jk} \ H_j(y) \int H_k(x) \ H_l(x) \ dx \\
	&=& \sum_{jk} P_{ik} \ L_{jk} \ H_j(y)
\eea

Start by writing the first order correction to the eigenvalue using this basis representation.
The first order correction to the eigenvalue is given in \refeq{EigVal}.
\bea
	\Gamma_i^{(1)}
	&=& \frac{\int\left(\sum_l \tilde{P}^{(0)}_{il} \ H_l(x)\right) \
	\left(\sum_{jk} P^{(0)}_{ik} \ L^{(1)}_{jk} \ H_j(x)\right) \ dx}
	{\int \left(\sum_j \tilde{P}^{(0)}_{ij} \ H_j(x)\right) \
	\left( \sum_k P^{(0)}_{ik} \ H_k(x) \right) \ dx} \\
	&=& \frac{\sum_{jk} \tilde{P}^{(0)}_{ij} \ P^{(0)}_{ik} \ L^{(1)}_{jk}}
	{\sum_j \tilde{P}^{(0)}_{ij} \ P^{(0)}_{ij}}
\eea
Write this in matrix notation.
\beq
	\Gamma_i^{(1)} = \frac{ \transp{\tilde{p}^{(0)}_i} \ \mathbf{L}^{(1)} \ p^{(0)}_i }
	{ \transp{\tilde{p}^{(0)}_i} \ p^{(0)}_i }
\eeq

To find the first order correction to the eigenfunction,
write \refeq{FirstOrderTerms} entirely in terms of this basis.
We are interested in finding the coefficients $P^{(1)}_{ik}$.

\beq
	\sum_{jk} P^{(1)}_{ik} \ L^{(0)}_{jk} \ H_j(x) +
	\sum_{mn} P^{(0)}_{in} \ L^{(1)}_{mn} \ H_m (x) =
	\Gamma^{(0)}_i \sum_a P^{(1)}_{ia} \ H_a (x) +
	\Gamma^{(1)}_i \sum_b P^{(0)}_{ib} \ H_b (x)
\eeq
Multiply this by $H_l (x)$ and integrate over $x$ to get an expression
relating only the coefficients.
\beq
	\sum_k P^{(1)}_{ik} \ L^{(0)}_{lk} + \sum_n P^{(0)}_{in} \ L^{(1)}_{ln}
	= \Gamma^{(0)}_i \ P^{(1)}_{il} + \Gamma^{(1)}_i \ P^{(0)}_{il}
\eeq
Rearrange this to make it look like eigenvalue equations.
This will help us isolate $P^{(1)}_{il}$.
\beq
	\sum_k \left( L^{(0)}_{lk} - \Gamma^{(0)}_i \delta_{lk} \right) \ P^{(1)}_{ik} =
	- \sum_n \left( L^{(1)}_{ln} - \Gamma^{(1)}_i \delta_{ln} \right) \ P^{(0)}_{in}
\eeq
Rewrite this in matrix notation.
\beq
	\left( \mathbf{L}^{(0)} - \Gamma^{(0)}_i \matId \right) p^{(1)}_i =
	- \left( \mathbf{L}^{(1)} - \Gamma^{(1)}_i \matId \right) p^{(0)}_i
\ee{Matrix1}
The matrix $\mathbf{L}^{(0)} - \Gamma^{(0)}_i \matId$ is not invertible
since $\Gamma^{(0)}_i$ is an eigenvalue of $\Lnoise{(0)}$.

We can use the pseudoinverse, now applied to a matrix,
to write an explicit expression for the coefficients
of the first order correction to the eigenfunction.
\beq
	p^{(1)}_i = - \left( \mathbf{L}^{(0)} - \Gamma^{(0)}_i \matId \right)^+
	\left( \mathbf{L}^{(1)} - \Gamma^{(1)}_i \matId \right) p^{(0)}_i .	
\ee{Final}

\section{The Moore-Penrose and other pseudoinverses}
\label{MPvsOth}

In this section we state the properties of the Moore-Penrose
pseudoinverse, the Drazin pseudoinverse, and the group inverse,
and show that they all coincide for the operator $\Lnoise{(0)}~-~\Gamma_i^{(0)}~\matId$,
used in the text.

The Moore-Penrose pseudoinverse of a rectangular matrix $A$ has the following properties~\rf{BarHuss11}
\bea
AA^+A = A, \\
A^+AA^+ = A^+, \\
\left(AA^+\right)^\dagger = AA^+, \\
\left(A^+A\right)^\dagger = A^+A .
\label{MPProps}
\eea
This operator is of common use, due to its generality.
A similar operation is performed by the
 Drazin (pseudo)inverse~\rf{Drazin58}, used for example by Kato~\rf{Kato80},
who calls it reduced resolvent (denoted as $\Lnoise{(0)}~-~\Gamma_i^{(0)}~\matId$, \rf{AvrFiHow13})
of the unperturbed operator $\Lnoise{(0)}$.
The Drazin inverse
is defined for a square matrix $A$ in an associative ring (or a semigroup),
and it has the following
properties:
\bea
A^{k+1}A^\# = A^k, \\
A^\#AA^\# = A^\#, \\
AA^\#A = A, \\
AA^\# = A^\#A.
\label{DrazProps}
\eea
Here $k=ind(A)$ is the index of $A$, that is the smallest nonnegative integer
such that $rank(A^k)=rank(A^{k+1})$. The index also coincides with the multiplicity
of the eigenvalue $0$, or equivalently, the dimension of the kernel of the matrix~\rf{AgCheb02}.
Importantly, Drazin and Moore-Penrose operators coincide and are called group inverse
if $ind(A)=~1$~\rf{BuFengDong12}.
That is the case of the operator  $\Lnoise{(0)}~-~\Gamma_i^{(0)}~\matId$,
whose kernel is one-dimensional since the eigenvalue $\Gamma_i^{(0)}$
of $\Lnoise{(0)}$ is simple.

For completeness, we include the properties of the group inverse~\rf{BenGre74,CamMey79}
\bea
AA^gA = A \\
A^gAA^g = A^g \\
AA^g = A^gA .
\label{group_inv}
\eea

\section*{References}

\bibliographystyle{model3-num-names}
\bibliography{../../bibtex/lippolis}

\ifboyscout
\newpage
\input{../../blog/flotperturb}
\fi

\end{document}